\documentclass[aps,prl,twocolumn,superscriptaddress]{revtex4}
\usepackage{amsmath}
\usepackage{amssymb}
\usepackage{graphicx}

\newcommand{\m}{{\bf m}}

\begin{document}

\title{
Spin-torque shot noise in magnetic tunnel junctions}

\author{A. L. Chudnovskiy}
\affiliation{1. Institut f\"ur Theoretische Physik, Universit\"at Hamburg,
Jungiusstr 9, D-20355 Hamburg, Germany}

\author{J. Swiebodzinski}
\affiliation{1. Institut f\"ur Theoretische Physik, Universit\"at Hamburg,
Jungiusstr 9, D-20355 Hamburg, Germany}

\author{A. Kamenev}
\affiliation{Department of Physics, University of Minnesota, Minneapolis,
Minnesota 55455, USA}

\date{\today}

\begin{abstract}
Spin polarized current may transfer angular momentum to a ferromagnet, resulting in 
a spin-torque phenomenon. At the same time the shot noise, associated with the current,
leads to a non-equilibrium stochastic force acting on the ferromagnet. We derive
stochastic version of Landau-Lifshitz-Gilbert equation for a magnetization of a ''free''
ferromagnetic layer in contact with a ''fixed'' ferromagnet. We solve the corresponding Fokker-Planck equation and
show that the non-equilibrium noise  yields  to a non-monotonous dependence of the precession
spectrum linewidth on the current.
\end{abstract}


\maketitle

Magnetization dynamics of a ferromagnet under influence of a
spin polarized current is a subject of intensive investigations (for recent reviews see Refs.~
\cite{Ralph-Stiles,Halperin-Tserkovnyak}).
It was realized \cite{Slonczewski, Berger} that the spin current may transfer the
angular momentum to the ferromagnet,  resulting in a torque acting on its magnetization
direction. In the case of a small ferromagnetic domain  the torque may lead to a
rotation of the magnetization  as a whole, rather than to an excitation of  spin waves.
This phenomenon, allowing for an electronic manipulation of the magnetization, has a promise
for  a number of  potential applications.

The effect has been recently observed \cite{Kiselev03,Rippard04,Ralph05,Rippard06,Mistral06} in a setup, where the spin-torque
magnitude and direction are tuned to compensate exactly the dissipation force acting on
the magnetization of the ''free'' ferromagnetic layer. This leads to an undamped precession
which is detected through the induced microwave radiation. Both the spectral width
and the generated power exhibit a strong dependence  on the current flowing through the interface of the
two ferromagnets.
It was shown later  \cite{Slavin07,Kim07} that the equilibrium thermal noise, first introduced
in dynamics of micromagnets   by F-L. Brown \cite{Brown63}, may partially account
for the observed linewidth.

On the other hand, since the experiments are performed under non-equilibrium conditions (spin current
strong enough to balance the dissipation), one needs to address non-thermal sources of noise as well.
The most essential of them is the {\em spin shot noise} associated with the discreteness of spin passing through the interface.  This phenomenon may be accounted for  by adding  a fluctuating part to the spin current {\em vector}
in the Slonczewski's torque term of Landau-Lifshitz-Gilbert (LLG) equation ${\bf I}_s\to {\bf I}_s + {\bf \delta I}_s(t)$.  The resulting {\em stochastic} LLG equation for the unit
vector ${\bf m}={\bf M}/M$ in the direction of the magnetization ${\bf M}$ takes the form
\begin{eqnarray}
\frac{d\m}{dt} &=& -\gamma \left[\m\times {\bf H}_{\mathrm{eff}}\right]+ \alpha(\theta) \left[\m  \times \frac{d\m}{dt}\right]
\nonumber \\
&+&\frac{\gamma}{M{\cal V}}\, \big[{\bf m}\times \left[\left({\bf I_s}+{\bf \delta  I}_s\right)  \times {\bf m}\right]\big] \,. \label{LLG}
\end{eqnarray}
Here $\gamma$ is gyromagnetic ratio, $ {\bf H}_{\mathrm{eff}}=-\partial F/\partial {\bf M}$ is the effective magnetic field, which includes both an external field and magnetic anisotropy, and ${\cal V}$ is a volume of the free ferromagnet.   Gilbert damping $\alpha(\theta)$ is renormalized by the coupling to the fixed ferromagnet \cite{Tserkovnyak02,Halperin-Tserkovnyak} and is thus dependent on a relative orientation angle $\theta$ of the fixed and free ferromagnets.

One could expect that the fluctuating part of the spin current vector ${\bf \delta I}_s(t)$ is preferentially directed
along the spin polarization of the incoming electron flux, i.e. along ${\bf I}_s$.
This is {\em not} the case,
however, due to the quantum nature of the effect. Indeed, each spin-flip event transfers exactly one $\hbar$ unit of the angular momentum to  the free ferromagnet. Due to the uncertainty principle,  direction
of an ensuing magnetization rotation is completely random. As a result, the fluctuating part of the spin torque must have  an isotropic correlator
\begin{equation}\label{correlator}
    \left\langle \delta {\rm I}_{s,i} \delta {\rm I}_{s,j} \right\rangle =  2{\cal D}(\theta)\,\delta_{ij}\,\delta(t-t')\, ,
\end{equation}
where the variance ${\cal D}(\theta)$ does not depend on the cartesian indexes  $i,j=x,y,z$, but may depend on the mutual orientation of the two ferromagnets. Because of the isotropy  of the stochastic torque, it can be equally
well represented by a fluctuating magnetic field  $ {\bf H}_{\mathrm{eff}}\to  {\bf H}_{\mathrm{eff}} +{\bf h}(t)$,  instead of the fluctuating spin flux ${\bf \delta I}_s(t)$. In the latter case the correlator of the stochastic fields reads as $\langle {\rm h}_i(t) {\rm h}_j(t')\rangle = 2{\cal D} (M{\cal V})^{-2} \delta_{ij}\delta(t-t')$. This type of non-equilibrium noise was considered in
Ref.~\cite{Tserkovnyak05} in context of NFN structures. 

Here we consider a model of a magnetic tunnel junction (MTJ) Ref.~
\cite{Slonczewski-Sun}.  Magnetization dynamics of the free ferromagnet is described  using
Holstein-Primakoff (HP) parametrization \cite{Holstein-Primakoff} of its total spin operator 
by deriving  semiclassical  equations
of motions for HP bosons. We employ  Keldysh formalism to allow for
non-equilibrium conditions, i.e. voltage bias between the two ferromagnets
\cite{Keldysh,Kamenev05,MacDonald}.
After  integrating  out the fermionic degrees of freedom in the second order in both
tunneling and spin-flip processes, we obtain an effective action for HP bosons, which encapsulates
deterministic forces (external magnetic field and spin-torque) along with the stochastic
term. The latter contains  an information about both equilibrium and non-equilibrium noise
components.

The result of the program, outlined above, is the following noise correlator
\begin{equation}
{\cal D}(\theta) =  \frac{M {\cal V}}{\gamma}\,\alpha_0 T + \frac{\hbar}{2}\, {\rm I}_{sf}(\theta) \coth\left(\frac{eV}{2T}\right)\, ,
\label{DI}
\end{equation}
where $\alpha_0$ is a bare Gilbert damping of an isolated grain and $V$ is a voltage
bias between the two ferromagnets.  The non-equilibrium part of the
noise is proportional to the  spin-flip current ${\rm I}_{sf}(\theta)$. The latter counts the number of
{\em real} spin flip events (as opposed  to the spin current ${\rm I}_s$ which is associated
with virtual spin flips).  In the MTJ setup we found for the spin-flip conductance
\begin{equation}\label{I_sf}
    \frac{d {\rm I}_{sf}(\theta)}{d V}  = \frac{\hbar}{4e} \left[
    G_{P} \sin^2\!\left(\frac{\theta}{2}\right)+ G_{AP} \cos^2\!\left( 
    \frac{\theta}{2}\right)\right] ;
\end{equation}
$$
G_P=G_{++}+G_{--}\,; \quad G_{AP} =G_{+-}+G_{-+}\, ,
$$
where we adopted notations of Ref.~\cite{Slonczewski-Sun} for the partial conductances $G_{\sigma\sigma'}$  between
the spin-polarized bands of the two ferromagnets. Here $G_{P/AP}$ stay for electric conductances in parallel (P) and antiparallel (AP) configurations, with $G_P\geq G_{AP}$. The electric conductance of the MTJ in an arbitrary orientation is given by
 $d{\rm I}_e(\theta)/dV=G_{P}\cos^2(\theta/2) + G_{AP}\sin^2(\theta/2)$. Notice that the spin shot noise is minimal for P  orientation and maximal for AP one -- exactly opposite to the charge current and the charge shot noise. In the same notations the spin conductance is \cite{Slonczewski-Sun}
\begin{equation}\label{I_s}
   \frac{ d{\rm I}_s}{dV} = \frac{\hbar}{4e}\left( G_{++}-G_{-+}+G_{+-}-G_{--}\right)\, ,
\end{equation}
where the angular dependence is already explicitly taken into account in Eq.~(\ref{LLG}).
A non-polarized current, i.e. $G_{+\sigma}=G_{-\sigma}$, does not exert deterministic spin torque, nevertheless it
still induces the shot noise torque on the ferromagnet. This effect was recently
discussed in the context of NFN structures \cite{Tserkovnyak05}.

The same calculation  also leads to a renormalization of Gilbert damping
coefficient in  LLG equation (\ref{LLG})
\begin{equation}\label{alpha-renorm}
    \alpha(\theta) = \alpha_0 +\frac{ \hbar \gamma}{ e M{\cal V}}
   \left( \frac{d {\rm I}_{sf}(\theta)}{d V} \right)\, ,
\end{equation}
where the spin-flip conductance is given by Eq.~(\ref{I_sf}).
Such an enhancement of the dissipation due to the spin transport between the two ferromagnets is
discussed in  Refs.~ \cite{Tserkovnyak02,Halperin-Tserkovnyak}.
In compliance with the fluctuation--dissipation theorem, the equilibrium
$(V\to 0)$ noise correlator is given by ${\cal D}=TM{\cal V}\alpha(\theta)/\gamma$
\cite{Brown63}.
Due to the renormalization, the damping is minimal in  P orientation.
The angular dependence of the damping term  selects  a unique angle, where
the spin-torque compensates dissipation. Upon increasing the spin current above the critical
one, such an angle gradually increases from zero to $\pi$.

Below we restrict ourselves to a setup, where both the magnetic field and the spin current are directed along the
$z$-axis specified  by  the magnetization direction  of the fixed
ferromagnet.   Transforming Eq. (\ref{LLG}) to the spherical coordinates \cite{vankampen}, one obtains two coupled  Langevin equations for the relative angle  $\theta(t)$ and the azimuthal angle $\phi(t)$ 
\begin{eqnarray}
\dot{\theta}&=& - \sin\theta \,T_z(\theta)  + \cot   \theta \, \tilde{\cal D}(\theta)
+   \xi_{\theta}(t)\, , \label{dottheta}\\
\dot{\phi}&=&  \Omega(\theta)
+  \csc \theta \, \xi_{\phi}(t)\, ;
\label{dotphi}
\end{eqnarray}
$$ T_z(\theta) = \gamma'\left(\alpha(\theta) { H}_{z}+ {{\rm I_s}/( M{\cal V})}\right)\,;\quad
\Omega  = \gamma H_z -\alpha T_z\,,
$$
where $\gamma'=\gamma/(1+\alpha^2)$ and $ \tilde{\cal D}(\theta)={\cal D}(\theta)\gamma\gamma'/( M{\cal V})^2$.
Here $\xi_{\theta}(t)$ and $\xi_{\phi}(t)$ are  two {\em uncorrelated} random noises  with the correlators
\begin{equation}
\label{angle-xi}
\langle\xi_\theta(t)\xi_\theta(t')\rangle=\langle\xi_\phi(t)\xi_\phi(t')\rangle=2\tilde{\cal D}(\theta)\delta(t-t')\, .
\end{equation}
Both Eq.~(\ref{LLG}) and Eqs.~(\ref{dottheta}), (\ref{dotphi}) should be interpreted in the sense of
retarded regularization, or Ito calculus \cite{vankampen}.

Let us first analyze  deterministic dynamics described by Eqs. (\ref{dottheta}),
(\ref{dotphi}) with $\tilde{\cal D}\to 0$. For a strong enough (and negative for positive $H_z$) spin current  the condition $T_z(\theta)=0$ may be  satisfied for a certain angle $\bar\theta$. Notice that it is the  angular dependence of the enhanced Gilbert damping \cite{Tserkovnyak02,Halperin-Tserkovnyak}, which is responsible for the angle selectivity.   In such a case Eq.~(\ref{dotphi}) describes a stable undamped precession with the frequency $\Omega(\bar\theta)=\gamma H_z$. The intensity of the induced microwave radiation \cite{Kiselev03,Rippard04,Ralph05,Rippard06,Mistral06} is given by the square of the oscillating magnetic moment, i.e. proportional to $\sin^2\bar\theta$.

To analyze effects of the noise we shall assume that $\alpha(\bar\theta) \ll 1$, allowing for
time scales separation. The fast variable is the azimuthal  angle $\phi(t)$, while the  angle
$\theta(t)$ is the slow one. For a fixed slow variable $\theta$ Eqs.~(\ref{dotphi}), (\ref{angle-xi}) lead to the   Lorentzian shape of the emitted microwave power spectrum
\begin{equation}\label{S}
    S(\omega,\theta) \propto  \sin^2\theta \,\frac{2 \csc^2 \theta\, \tilde{\cal D}(\theta)  }
    {\left[\omega-\Omega(\theta)\right]^2 +
    \left[\csc^2 \theta\, \tilde{\cal D}(\theta) \right]^2 } \,.
\end{equation}
To compare with the observed power spectrum  this expression should be  averaged over the stationary probability distribution $P(\theta)$ of the slow degree of freedom
$S(\omega) = \int \sin\theta d\theta\, P(\theta) S(\omega,\theta)$.
The distribution function $P(\theta,t)$ obeys the Fokker-Planck equation which follows from Eqs.~(\ref{dottheta}), (\ref{angle-xi}) \cite{Brown63,vankampen}
\begin{eqnarray}
\dot{P}=\frac{1}{\sin\theta}\, \partial_\theta\left[\sin^2\theta\,  T_z(\theta)  P +\sin\theta\, \partial_\theta \left( \tilde{\cal D}(\theta) P \right) \right].
\label{FP}
\end{eqnarray}
The stationary solution of Eq.~(\ref{FP})
is given by
\begin{equation}
\label{W}
P(\theta)= \frac{1}{Z \tilde{\cal D}(\theta)} \, \exp\left\{ - \int_0^\theta \sin\theta'\, d\theta' \,
\frac{T_z(\theta')}{\tilde{\cal D}(\theta')} \right\},
\end{equation}
where constant $Z$ is chosen to satisfy the normalization condition  $\int_0^\pi \sin\theta\, d\theta\, P(\theta)=1$. For a weak noise the stationary distribution function has a sharp maximum close to the angle $\bar\theta$, where the deterministic spin-torque compensates the dissipation.

\begin{figure}
\includegraphics[width=8cm,height=7cm,angle=0]{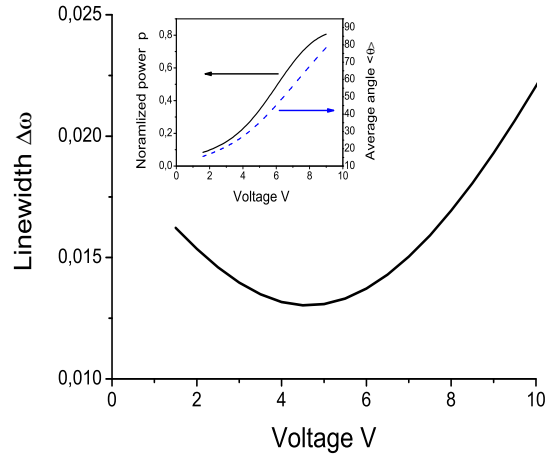}%
\caption{(Color online) Calculated linewidth of the microwave spectral
power vs. applied voltage in units of temperature.
Inset: calculated microwave power (solid line) and the average
precession angle  (dashed line) vs. voltage.
Parameters: $M{\cal V}/(\hbar \gamma)=10$, $\hbar\gamma H_z=3$K, $T=1$K, 
$\alpha_0=0.01$.  Conductances in units $e^2/h$: $G_P=0.181$, $G_{AP}=0.019$. 
$dI_s/dV=0.01 e$. 
\label{fig1}}
\end{figure}

Figure \ref{fig1} shows the calculated spectral linewidth at the half maximum as a function of the applied voltage.
The origin of the non-monotonous dependence may be understood by inspection of   Eq.~(\ref{S}). The initial decline is due to the geometric factor $\csc^2 \theta$ in the width of the Lorentzian. As the voltage increases, so does the
angle where the distribution function exhibits the maximum. Due to $\csc\theta$, coming from the noise term on the r.h.s. of Eq.~(\ref{dotphi}), the amplitude of the phase noise decreases leading to a narrowing of the power spectrum. The initial decrease of the spectral width was discussed in Ref.~\cite{Kim07,Slavin07} in terms of the equilibrium thermal noise \cite{Brown63}.

The subsequent increase of the spectral width at yet larger voltages was observed in a number of experiments
\cite{Rippard04,Rippard06,Mistral06}  and, to the best of our knowledge, remained unexplained. It naturally comes about due to
the non-equilibrium component of the noise. Indeed the noise correlator (\ref{DI}) grows with the applied
voltage due to the growth of the spin-flip current. The dependence of the spin-flip current ${\rm I}_{sf}(\theta)$ on the bias is faster than the linear because of 
the increase of the spin-flip conductance, Eq.~(\ref{I_sf}),
with the operation angle $\bar\theta$ on top of the overall proportionality of ${\rm I}_{sf}$ to the bias $V$.
As a result, for  $eV\gtrsim 2T$ the noise variance ${\cal D}$  grows rapidly, leading to the broadening of the power
spectrum, cf.~Eq.~(\ref{S}).

Another consequence of the non-equilibrium noise is the saturation of the spectral linewidth at small temperatures \cite{Ralph05}. Since the devices are always operated at currents larger than the critical one, the noise intensity (\ref{DI}) is finite ${\cal D}(\theta)=(\hbar/2) {\rm I}_{sf}(\theta)$ even at $T\to 0$. Thus
decreasing the temperature one should observe saturation of the linewidth at $T\sim eV$, provided the induced
damping (the second term on the r.h.s. of Eq.~(\ref{alpha-renorm})) is larger than the bare one.

In the remainder of the paper we outline  derivation of Eqs.~(\ref{LLG}) -- (\ref{alpha-renorm}).
The MTJ is modeled by the two itinerant ferromagnets, whose majority ($\sigma=+$) and minority ($\sigma=-$) bands
are described by the operators $c^{\dagger}_{k\sigma}, c_{k\sigma}$ for the fixed ferromagnet and
$ d^\dagger_{l\sigma}, d_{l\sigma}$ for the free layer. We found convenient  to work in the instantaneous reference frame, where the magnetization of the free layer points in the $z$-direction.   The corresponding Hamiltonian takes the following form
\begin{eqnarray}\label{Ham2}
\nonumber &&
    H_0=\sum_{k,\sigma} \epsilon_{k\sigma} c^{\dagger}_{k\sigma} c_{k\sigma} +
    \sum_{l\sigma}(\epsilon_{l} -  J S_z\sigma) d^\dagger_{l\sigma} d_{l\sigma} -\gamma  {\bf S}\cdot {\bf H}
  \\
\nonumber &&
- J\,( { S_+}{s_-}
    +  { S_-} { s_+})+
    \left[
    \sum_{kl,\sigma\sigma'}
W_{kl}^{\sigma\sigma'} c^\dagger_{k\sigma} d_{l\sigma'} + h.c.
\right] \,,
\end{eqnarray}
 here the spin-dependent tunneling matrix elements are $W_{kl}^{\sigma\sigma'} =  \langle \sigma|\sigma'\rangle W$, where the spin-transformation matrix is $\langle \sigma|\sigma\rangle=
e^{-i\sigma\phi/2} \cos\theta/2\,$ and $\langle \sigma|\sigma'\rangle=
e^{i\sigma\phi/2} \sin\theta/2$;  and ${\bf s}=\frac{1}{2}\sum_{l\sigma\sigma'} d^\dagger_{l\sigma} \vec\sigma_{\sigma\sigma'} d_{l\sigma'}$ is the spin of itinerant electrons, while $s_\pm=s_x\pm i s_y$.
We have explicitly accounted for the interactions of the itinerant electrons in the free layer with its {\em total} spin ${\bf S} = {\bf M}{\cal V}/\gamma$. To make the latter a dynamical variable we use HP parametrization \cite{Holstein-Primakoff}
$$ S_z=S- b^\dagger b;   \  \
S_-= b^\dagger\sqrt{2S-b^\dagger b};   \  \
S_+= \sqrt{2S-b^\dagger b} \,b, $$
where $b^\dagger, b$ are usual bosonic operators.

\begin{figure}
\includegraphics[width=8cm,height=3cm,angle=0]{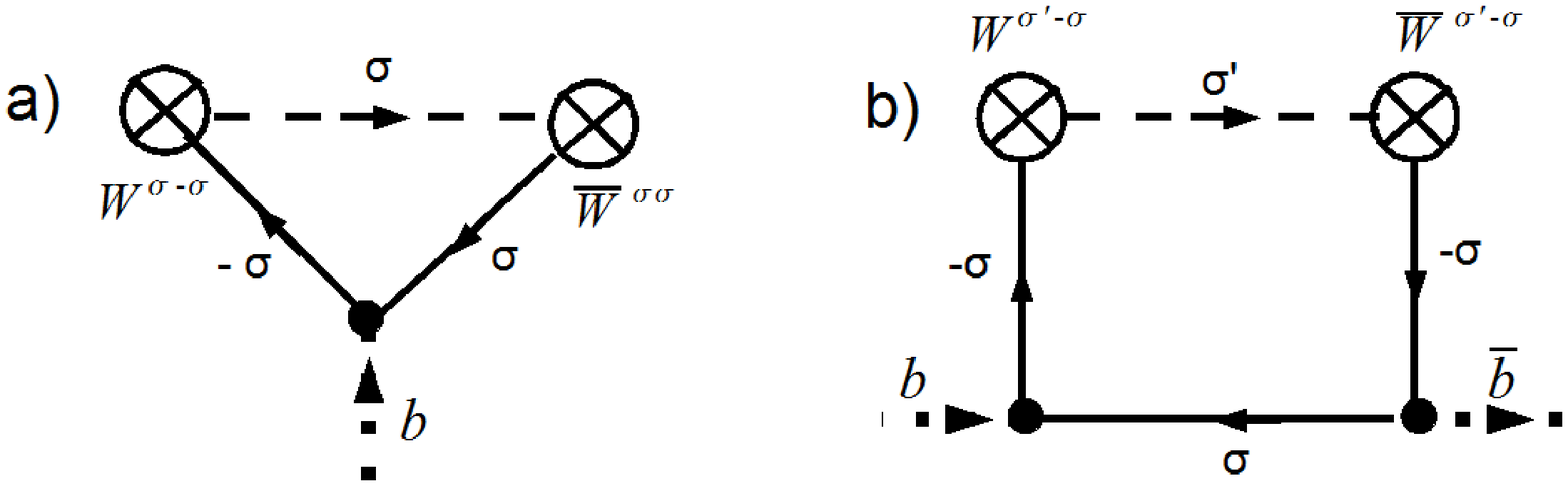}%
\caption{Diagrams for spin-flip processes:  (a) the first order, describing the spin-torque;
(b) the second order, describing the spin shot
noise along  with the enhanced   damping.
Solid (dashed) lines denote electronic propagators in the free (fixed)  layers.
Bold dashed lines are  propagators of HP  bosons (spin-flips).
Tunneling vertices are denoted by circles with crosses.
\label{fig2}}
\end{figure}

Next we  write the corresponding action in terms of complex fermionic and bosonic fields $c_{k\sigma}(t), d_{l\sigma}(t), b(t)$, where the time variable runs along the closed Keldysh contour \cite{Keldysh,Kamenev05,MacDonald}. We then transform  to two-component vector notations in terms of
symmetric (classical "$cl$") and antisymmetric (quantum "$q$") combinations of the forward and backward propagating fields. One should keep in mind that the distribution functions of the $c$ and $d$ fermions have a relative shift of the chemical potentials  by $eV$. The fermionic fields may be integrated out exactly and the remaining bosonic  effective action expanded to the second order in the tunneling amplitude $W$ and to the first and second orders in the spin-flip processes $S_\pm s_\mp$.  The corresponding processes are represented by  the diagrams of Fig.~\ref{fig2}. The approximations are justified by the weakness of tunneling and largeness of $S\gg \hbar $.

The resulting action for the complex bosonic fields $b_{cl}(t), b_q(t)$ takes the form
${\cal S}= {\cal S}_0+{\cal S}_1+{\cal S}_2$ where the subscript indicates the order in spin-flips processes. Here the bare action is \cite{foot}
\begin{equation}\label{S0}
    {\cal S}_0=\int\!\! dt\,  \bar b_q(t)\left( i\partial_t b_{cl}(t) +\gamma\sqrt{S/2}\, H_+\right)+ c.c. \,
\end{equation}
The first order correction in spin-flip amplitude is represented by diagram of Fig.~\ref{fig2}a. This is
a virtual transition into an opposite spin band with a subsequent tunneling out of the free layer into the ''correct'' spin band of the fixed magnet.
The latter process is possible for $\theta\neq 0,\pi$, due to a finite $W^{\sigma\sigma'}_{kl}$. The net result is
transferring angular momentum $\hbar$ to the total spin of the free layer, i.e. the deterministic spin-torque \cite{Slonczewski,Berger}. The corresponding contribution to the action is
\begin{equation}
{\cal S}_{1}=\frac{i}{\sqrt{2S}}\int\!\! dt\, \bar b_q(t){\rm I}_s \sin\theta\, e^{-i\phi} +c.c. \, ,
\label{deltaS1}
\end{equation}
where ${\rm I}_s$ is given by Eq.~(\ref{I_s}) with $G_{\sigma\sigma'} = 
\frac{4\pi e^2}{h} |W|^2\nu_c^\sigma \nu_d^{\sigma'}$ and
$\nu_{c,d}^\sigma$ are densities of states of the two ferromagnets in the $\sigma$ band.   The second order processes in spin-flips are depicted by the diagram of Fig.~\ref{fig2}b. These are  real (i.e. Golden rule) processes, which  matrix elements include the spin-flips. They lead to dissipation as well as fluctuations.
The  corresponding action is
\begin{eqnarray}
{\cal S}_{2} = \int\!\! dt\, \left[\alpha(\theta) (\bar b_q\partial_t b_{cl} -
\bar b_{cl}\partial_t  b_{q}) + \frac{2i}{S}\, {\cal D}(\theta) \bar b_q b_q \right],
 \label{deltaS2}
\end{eqnarray}
where ${\cal D}(\theta)$ and $\alpha(\theta)$  are given by Eqs.~(\ref{DI}), (\ref{I_sf}) and (\ref{alpha-renorm}) (without internal dissipation $\alpha_0$).

One then decouples the last term on the r.h.s. of Eq.~(\ref{deltaS2}) by means of the complex Hubbard-Stratonovich
field $\delta {\rm I}_+(t)={\rm I}_{s,x}+i{\rm I}_{s,y} $. The remaining action is linear in $b_q(t)$ and $\bar b_q(t)$. It constitutes thus  resolution of functional $\delta$-functions of the first order Langevin equations on
$b_{cl}(t)=\sqrt{M{\cal V}/(2\gamma})\, m_+(t)$ and its complex conjugate. Those are nothing but $m_\pm$ components of Eq.~(\ref{LLG}), with the noise intensity given by Eqs.~(\ref{correlator})--(\ref{I_sf}) \cite{foot}.

To conclude: in the presence of the spin-torque, caused by a spin polarized current,
the LLG equation should be modified to include a stochastic Langevin term which accounts for the shot-noise, associated
with the spin current. This term is different from the previously discussed thermal stochasticity in LLG equation
\cite{Brown63}, because of its non-equilibrium origin. We have derived the corresponding noise correlator
in the MTJ setup.   We have argued that the non-equilibrium noise manifests itself in a non-monotonous
voltage dependence and low-temperature saturation of the linewidth of the precession power spectrum.

\begin{acknowledgments}

We are grateful to P.~Crowell, A.~Levchenko, D. Pfannkuche and V. Kagalovsky 
for useful discussions.
A. C. and J. S.  acknowledge financial support from DFG through
Sonderforschungsbereich 508 and Sonderforschungsbereich 668. A.K. was supported
by the NSF grant  DMR-0405212  and by the A.~P.~Sloan foundation.

\end{acknowledgments}


\begin{thebibliography}{99}
\bibitem{foot} In the instanteneous reference frame  $m_z=1$ and $m_\pm=0$, while $\dot m_z=0$ and $\dot m_\pm\neq 0$.
Therefore all the terms in the action  $\sim b_{cl}(t)\propto m_+(t)$ do not contribute to the equations of motion and are omitted. We thus keep only terms $\sim \partial_t b_{cl}$ along with those which contain only $b_q$.
\end{thebibliography}
\end{document}